\documentclass[conference]{IEEEtran} 
\usepackage[ruled,linesnumbered]{algorithm2e}
\makeatletter
\newcommand{\RemoveAlgoNumber}{\renewcommand{\fnum@algocf}{\AlCapSty{\AlCapFnt\algorithmcfname}}}
\newcommand{\RevertAlgoNumber}{\algocf@resetfnum}
\makeatother

\IEEEoverridecommandlockouts

\usepackage{cite}
\usepackage{amsmath, amssymb, mathtools}
\usepackage[mathscr]{eucal}
\usepackage{algorithmic}
\usepackage{graphicx}
\usepackage{textcomp}
\usepackage{xcolor}
\usepackage{booktabs}
\usepackage[Symbol]{upgreek}

\usepackage{geometry} 				
\def\BibTeX{{\rm B\kern-.05em{\sc i\kern-.025em b}\kern-.08em
    T\kern-.1667em\lower.7ex\hbox{E}\kern-.125emX}}

\graphicspath{{./Figures/}}

\newtheorem{Lemma}{Lemma}
\newtheorem{Theorem}{Theorem}
\newtheorem{Corollary}{Corollary}

\newcommand{\uth}{\underline{\textnormal{th}}}
\newcommand{\tsum}{\operatorname{\textstyle{\sum}}}

\begin{document}
\setlength{\textfloatsep}{0pt}
\setlength{\abovecaptionskip}{-5pt}
\setlength{\belowcaptionskip}{-10pt}
\setlength\intextsep{0pt}

\title{Beyond Diagonal RIS for ISAC Network: Statistical Analysis and Network Parameter Estimation}

\author{\IEEEauthorblockN{
        Thanh~Luan~Nguyen{\IEEEauthorrefmark{1}}, 
        Georges~Kaddoum{\IEEEauthorrefmark{1}},
        Bassant Selim{\IEEEauthorrefmark{1}},
        Chadi Assi{\IEEEauthorrefmark{2}},
        }
    \IEEEauthorrefmark{1} Department of Electrical Engineering, \'{E}cole de Technologie Sup\'{e}rieure (\'{E}TS), Montr\'{e}al, QC, Canada \\
    \IEEEauthorrefmark{2} Concordia Institute for Information Systems Engineering \\
      Emails: 
        thanh-luan.nguyen.1@ens.etsmtl.ca,
        georges.kaddoum@etsmtl.ca, \\
        bassant.selim@etsmtl.ca,
        chadi.assi@concordia.ca
}

\maketitle

\begin{abstract}
This paper investigates the use of beyond diagonal reconfigurable intelligent surface (BD-RIS) with $N$ elements to advance integrated sensing and communication (ISAC). 
We address a key gap in the statistical characterizations of the radar signal-to-noise ratio (SNR) and the communication signal-to-interference-plus-noise ratio (SINR) by deriving tractable closed-form cumulative distribution functions (CDFs) for these metrics. Our approach maximizes the radar SNR by jointly configuring radar beamforming and BD-RIS phase shifts. Subsequently, zero-forcing is adopted to mitigate user interference, enhancing the communication SINR. To meet ISAC outage requirements, we propose an analytically-driven successive non-inversion sampling (SNIS) algorithm for estimating network parameters satisfying network outage constraints. Numerical results illustrate the accuracy of the derived CDFs and demonstrate the effectiveness of the proposed SNIS algorithm.
\end{abstract}

\begin{IEEEkeywords}
Integrated sensing and communication (ISAC), beyond diagonal reconfigurable intelligent surface (BD-RIS), inversion sampling.
\end{IEEEkeywords}

\section{Introduction}

\IEEEPARstart{I}{ntegrated} sensing and communication (ISAC) has emerged as a key technology for 6G wireless communication systems \cite{Liu2022JSAC}, \cite{Wei2023IoTJ}. 
    Moreover, the ever-growing and ever-changing advancements in reconfigurable intelligent surface (RIS) technologies further unlock ISAC’s untapped potential for achieving more efficient and reliable systems \cite{Luo2023TVT, Yang2024TWC, Liu2024TWC, Hua2024TWC}. 
    Notably, \cite{Yang2024TWC} highlighted that RIS can boost both communication data rates and detection probabilities in multicell ISAC scenarios. 
    Furthermore, the work in \cite{Liu2024TWC} showed that optimized RIS configurations can remarkably improve communication rates and target detection, particularly through enhancing radar signal-to-noise ratio (SNR).

With the rapid development of the RIS technologies, a new beyond diagonal (BD)-RIS architecture has recently emerged and is often referred to as the “RIS 2.0” \cite{Li2023TWC, Li2024CM}. 
    Unlike diagonal RISs, the BD-RIS enhances functionality by linking different RIS elements with adjustable impedance components, which significantly improves its beam-steering capabilities. 
    A notable BD-RIS configuration is the fully-connected (FC) architecture \cite{Li2024CM, Liu2024WCL, Guang2024TVT}, where each pair of RIS elements is interconnected to maximize beamforming gain.
According to recent studies \cite{Li2023TWC, Liu2024WCL, Guang2024TVT}, the FC BD-RIS can enhance system performance more effectively than traditional diagonal RISs. 
    For example, \cite{Li2023TWC} investigated BD-RIS applications to improve spectral efficiency, showing that BD-RIS provides higher beamforming gains than its diagonal version. 
    Additionally, \cite{Liu2024WCL} found that FC BD-RIS can achieve up to 15\% increase in sum-rate compared to diagonal alternatives.
    Moreover, \cite{Guang2024TVT} demonstrated that the FC BD-RIS lowers transmit power consumption by 25\%–30\% compared to its diagonal counterpart.
    
While the use of FC BD-RIS in ISAC systems has received increasing attention in the literature, a significant theoretical gap remains in the statistical characterization of radar SNR and communication signal-to-interference-plus-noise ratio (SINR). 
    To address this gap, we aim to develop tractable, closed-form expressions for their cumulative distribution functions (CDFs), which are fundamental to many key design metrics, such as the outage probability.
Building on the derived expressions, we propose a novel algorithm for estimating network parameters to meet ISAC outage requirements.
The contributions of this paper are listed as follows:

\begin{itemize}
    \item This paper investigates using FC BD-RIS to enhance the performance of ISAC networks. Specifically, we prioritize maximizing the radar SNR by jointly optimizing the radar beamforming vector and the phase-shift matrix of the FC BD-RIS. 
    Next, we implement zero-forcing (ZF) beamforming to effectively mitigate total interference at each user, thereby enhancing communication SINR.
    This two-pronged approach allows us to derive tractable expressions for both radar SNR and communication SINR.
    \item Using the derived expressions, we obtain accurate and tractable closed-form CDFs for the radar SNR and communication SINR, particularly when the number of RIS elements is sufficiently large compared to the number of antennas at the base station (BS).
    \item We propose a novel analytically-driven successive non-inversion sampling (SNIS) algorithm designed for network parameter estimation under maximum network outage equality constraints. This algorithm can also be adopted to address inverse problems, such as parameter estimation and root-finding, within given constraints. Additionally, numerical alternatives to the SNIS algorithm are provided.
    \item Numerical results are provided to validate the accuracy of the derived CDFs and demonstrate the effectiveness of our proposed algorithm.
\end{itemize}

{\textit{Notation}}: $\mathbb{E}[\cdot]$ and $\mathbb{V}[\cdot]$ denote the expectation and variance operators, respectively; 
    $X \mathop{=}^d Y$ means that the random variables (RVs) $X \mathop{\to}^d Y$ have the same distribution, while $X \to Y$ means that $X$ converges to $Y$ in distribution; 
    $\mathbf{x}_{(l)}$ and $\mathbf{x}^{(L-l)}$ denote the first $l$ elements and the last $L-l$ elements of a $L$-dimensional vector $\mathbf{x}$, respectively.

\section{System Model}

We consider a multi-user ISAC system aided by a FC BD-RIS \cite{Liu2024WCL}. 
    The dual-functional BS is equipped with a uniform linear array (ULA) with $M$ antennas for transmitting and receiving. 
Moreover, the BS is simultaneously serving $K$ single-antenna mobile users ($K \le M$) and sensing a target located in a specific location. 
    The transmit signal $\mathbf{x} \in \mathbb{C}^M$ at the BS is 
\begin{align}
\mathbf{x} 
    &=  \sum_{k=1}^{K} \sqrt{P_{{\tt c}, k}} \mathbf{w}_{{\tt c}, k} s_{{\tt c}, k}
    +   \sum_{m=1}^{M} \sqrt{P_{{\tt r}, m}} \mathbf{w}_{{\tt r}, m} s_{{\tt r}, m},
\end{align}
where ${\mathbf{w}_{{\tt c}, k} \!\in\! \mathbb{C}^{M}}$ and 
    ${\mathbf{w}_{{\tt r}, m} \!\in\! \mathbb{C}^{M}}$ are the beamforming vectors for the communication signal of user $k \!\in\! {\cal K} \!\triangleq\! \{1, 2, \dots, K\}$ and radar signal $m \!\in\! {\cal M} \!\triangleq\! \{1, 2, \dots, M\}$, respectively,
    $P_{{\tt c}, k}$ (W) and $P_{{\tt r}, m}$ (W) are the corresponding allocated powers in Watts, 
and $s_{{\tt c}, k}$ and $s_{{\tt r}, m}$ are the corresponding unit-energy communication and radar signals. 
We assume that the communication and radar signals are statistically independent \cite{Luo2023TVT, Liu2024WCL}, i.e., 
    $\mathbb{E}\{ s_{{\tt r}, m} s_{{\tt c},k}^{\textnormal{H}} \} \!=\! 0$, with
    $s_{{\tt c}, k} \!\sim\! {\cal CN}(0, 1)$ and 
    $s_{{\tt r}, m} \!\sim\! {\cal CN}(0, 1)$, $\forall k \in {\cal K}$ and $\forall m \in {\cal M}$. 
    Moreover, the channel coefficients from the BS to the RIS, RIS to user $k \in {\cal K}$, and RIS to target are denoted by 
    $\sqrt{\mathfrak{L} d^{-\alpha}} \mathbf{G}$, 
    $\sqrt{\mathfrak{L} d_{{\tt c}, k}^{-\alpha}} \mathbf{h}_{{\tt c}, k}$, and 
    $\sqrt{\mathfrak{L} d_{\tt r}^{-\alpha}} \mathbf{h}_{{\tt r}, t}$, respectively. 
    Here, $d$, $d_{{\tt c}, k}$, and $d_{\tt r}$ are the corresponding distances in meters with $\mathfrak{L}$ and $\alpha$ being the path loss at the reference distance and path loss exponent, respectively, and $\mathbf{G} \!\in\! \mathbb{C}^{N \times M}$, 
    $\mathbf{h}_{{\tt c}, k} \!\in\! \mathbb{C}^N$, and 
    $\mathbf{h}_{{\tt r}, t} \!\in\! \mathbb{C}^N$ denote the small-scale fading with statistically independent and identically distributed (i.i.d.) entries, which follow circularly complex Gaussian distributions.
    Furthermore, we assume that direct links from the BS to both the target and mobile users are unavailable, a situation commonly encountered in densely populated urban environments where obstacles, such as buildings, walls, and other structures, obstruct line-of-sight communications \cite{NOMA3GPP}.
    Due to its fully-connected impedance networking architecture \cite{Li2023TWC, Li2024CM}, the FC BD-RIS phase configuration matrix $\mathbf{\Theta} \in \mathbb{C}^{M\times M}$ should satisfy $\mathbf{\Theta} \mathbf{\Theta}^{\textnormal{H}} \preceq \mathbf{I}_N$. 
Accordingly, the received signal at user $k \in {\cal K}$ is obtained as
\begin{align}
y_{{\tt c}, k}
    &=  
    \underbrace{ \sqrt{\mathfrak{L}_{{\tt c}, k} P_{{\tt c}, k}} \mathbf{h}_k^{\textnormal{H}} \mathbf{w}_{{\tt c}, k} s_{{\tt c}, k}}\limits_{\textnormal{desired signal}}
    +   \underbrace{\tsum_{\substack{i=1\\i\ne k}}{\sqrt{\mathfrak{L}_{{\tt c}, k} P_{{\tt c}, i}} \mathbf{h}_k^{\textnormal{H}} \mathbf{w}_{{\tt c}, i} s_{{\tt c}, i}}}\limits_{\textnormal{inter-user interference}}
    \nonumber\\
    &\quad
    +   \underbrace{\tsum_{m = 1}^{M}{\sqrt{\mathfrak{L}_{{\tt c}, k} P_{{\tt r}, m}} \mathbf{h}_k^{\textnormal{H}} \mathbf{w}_{{\tt r}, m} s_{{\tt r}, m}}}\limits_{\textnormal{sensing interference}} 
    + \underbrace{n_k}\limits_{\textnormal{noise}}, 
\end{align}
where 
    $\mathfrak{L}_{{\tt c}, k} \triangleq \mathfrak{L} d^{-\alpha} \mathfrak{L} d_{{\tt c}, k}^{-\alpha}$ and 
    $\mathbf{h}_k^{\textnormal{H}} \triangleq \mathbf{h}_{{\tt c}, k}^{\textnormal{H}} \mathbf{\Theta} \mathbf{G} \in \mathbb{C}^{M}$ denote the effective path loss and the effective channel vector from BS to user $k \in {\cal K}$ through the BD-RIS, respectively, and $n_k$ is the additive white Gaussian noise at user $k \in {\cal K}$, which follows zero-mean circularly complex Gaussian distribution with variance $\sigma_k^2$~(W).
Hence, the received signal-to-interference-plus-noise ratio (SINR) at the user $k \in {\cal K}$~is given by
\begin{align}
\Gamma_{{\tt c}, k}
\!=\!   \frac{ \bar{\gamma}_{{\tt c}, k} \! \left\vert \mathbf{h}_k^\textnormal{H} \mathbf{w}_{{\tt r}, k} \right\vert^2 }
        {   
            \sum_{\substack{i=1\\i\ne k}}^{K}{ \bar{\gamma}_{{\tt c}, i} \left| \mathbf{h}_k^{\textnormal{H}} \mathbf{w}_{{\tt r}, i} \right|^2 } 
            \! + \mathfrak{L}_{{\tt c}, k} \sum_{m = 1}^{M}{ \! \frac{P_{{\tt r}, m}}{\sigma_k^2} \! \left| \mathbf{h}_k^{\textnormal{H}} \mathbf{w}_{{\tt r}, m} \right|^2 } \! + 1
        },
\label{eq:SINR1}
\end{align}
where 
    $\bar{\gamma}_{{\tt c}, k} \!\triangleq\! \mathfrak{L}_{{\tt c}, k} P_{{\tt c}, k} / \sigma_k^2$, $\forall k \!\in\! {\cal K}$, and
    $\bar{\gamma}_{{\tt c}, i} \!\triangleq\! \mathfrak{L}_{{\tt c}, k} P_{{\tt c}, i} / \sigma_k^2$, $\forall i \!\in\! {\cal K} \backslash \{k\}$.
Taking into account the reciprocity of the channel, the echo signal reflected back from the sensing target is 
    $\mathbf{y}_{\tt r} \!=\! \alpha_{\tt r} \mathfrak{L}_{\tt r} \left( \mathbf{h}_{{\tt r}, t}^{\textnormal{H}} \mathbf{\Theta} \mathbf{G} \right)^{\textnormal{H}} \left( \mathbf{h}_{{\tt r}, t}^{\textnormal{H}} \mathbf{\Theta} \mathbf{G} \right) \mathbf{x} + \mathbf{n}_{\tt r}$~\cite{Liu2024WCL}, where 
    $\mathfrak{L}_{\tt r} \!\triangleq\! \mathfrak{L} d^{-\alpha} \mathfrak{L} d_{\tt r}^{-\alpha}$ denotes the effective path loss from the BS to the target through the RIS, $\alpha_{\tt r}$ represents the reflection cross-section (RCS) coefficient with mean power $\mathbb{E}\{ |\alpha_{\tt r}|^2 \} \!=\! \varsigma_{\tt r}^2$ and the noise at the receiver is modeled as $\mathbf{n}_{\tt r} \!\sim\! {\cal CN}\left( \mathbf{0}, \sigma_{\tt r}^2 \mathbf{I}_M \right)$ \cite{Luo2023TVT, Liu2024WCL}.
    We consider perfect self-interference cancellation, where the communication signals are perfectly cancelled out from the echo, 
    the signal-to-noise ratio (SNR) of the filtered sensing signal, i.e., the radar SNR, is \cite{Luo2023TVT}
\begin{align}
\Gamma_{{\tt r}, t} \!=\!\!
    \sum_{m=1}^{M}\! 
    \frac{\varsigma_{\tt r}^2}{\sigma_{\tt r}^2} P_{{\tt r}, m}
    \left\|
        \mathfrak{L}_{\tt r}
        \left( \mathbf{h}_{{\tt r}, t}^{\textnormal{H}} \mathbf{\Theta} \mathbf{G} \right)^{\textnormal{H}} 
        \left( \mathbf{h}_{{\tt r}, t}^{\textnormal{H}} \mathbf{\Theta} \mathbf{G} \right) \mathbf{w}_{{\tt r}, m} 
    \right\|^2. \label{eq:Gamma_t} 
\end{align}

\subsection{Maximum Achievable Radar SNR}

In this subsection, we derive the expression for the maximum radar SNR by jointly optimizing the radar beamforming vector and the phase-shift matrix of the BD-RIS.

\begin{Theorem}
\label{theo:maxSNR}
The maximum achievable radar SNR with respect to the radar beamforming and the phase-shift matrix is expressed as
\begin{align}
\Gamma_{{\tt r}, t}^{\star}
    =   \frac{\bar{\gamma}_{{\tt r}, t}}{M} \left| \mathbf{h}_{{\tt r}, t}^{\textnormal{H}} \mathbf{\Sigma} \mathbf{h}_{{\tt r}, t} \right|^2,
\label{eq:maxSNR}
\end{align}
where $\mathbf{\Sigma} = \operatorname{diag}( \lambda_{\langle 1 \rangle}, \lambda_{\langle 2 \rangle}, \dots, \lambda_{\langle M \rangle}, \mathbf{0}_{N-M})$ 
    is a diagonal matrix with the first $M$ diagonal elements being the positive eigenvalues of the Hermitian matrix $\mathbf{G} \mathbf{G}^{\textnormal{H}}$ in descending order and the remaining elements being zeros, ${P_{\tt r} \!=\! \sum_{m=1}^{M}{P_{{\tt r}, m}}}$ (W) is the total power allocated to radar signals in Watts, and
    $\bar{\gamma}_{{\tt r}, t} \!\triangleq\! P_{\tt r} \mathfrak{L}_{\tt r}^2 \varsigma_{\tt r}^2 / \sigma_{\tt r}^2$.
\end{Theorem}
\begin{IEEEproof}
The radar SNR in \eqref{eq:Gamma_t} is rewritten as
\begin{align}
\Gamma_{{\tt r}, t}
    \!=\! \left\| \mathfrak{L}_{\tt r} \mathbf{h}_{{\tt r}, t}^{\textnormal{H}} \mathbf{\Theta} \mathbf{G} \right\|^2 \!
    \sum_{m=1}^{M} \frac{\varsigma_{\tt r}^2}{\sigma_{\tt r}^2} P_{{\tt r}, m} 
    \left| 
        \mathbf{h}_{{\tt r}, t}^{\textnormal{H}} \mathbf{\Theta} \mathbf{G}
        \mathbf{w}_{{\tt r}, m}
    \right|^2, \label{eq:Gamma_t2} 
\end{align}
which is maximized with respect to ${\mathbf{w}_{{\tt r}, m} \in \mathbb{C}^{M}}$ as
\begin{align}
\Gamma_{{\tt r}, t} \!=\! 
    \frac{\bar{\gamma}_{{\tt r}, t}}{M} 
    \left| 
        \mathbf{h}_{{\tt r}, t}^{\textnormal{H}} \mathbf{\Theta} \mathbf{G} \mathbf{G}^{\textnormal{H}} \mathbf{\Theta}^{\textnormal{H}} \mathbf{h}_{{\tt r}, t} 
    \right|^2.
\label{eq:radarSNR_MRT}
\end{align}

    Moreover, by performing singular value decomposition $\mathbf{G} \!=\! \mathbf{U} \mathbf{\Sigma}^{\frac{1}{2}} \mathbf{V}^{\textnormal{H}}$, where $\mathbf{U} \in \mathbb{C}^{N\times N}$ and $\mathbf{V} \in \mathbb{C}^{M\times M}$ are unitary matrices with ${\mathbf{U}^{\textnormal{H}} \mathbf{U} \!=\! \mathbf{I}_N}$ and ${\mathbf{V} \mathbf{V}^{\textnormal{H}} \!=\! \mathbf{I}_M}$, the quadratic form $\mathbf{h}_{{\tt r}, t}^{\textnormal{H}} \mathbf{\Theta} \mathbf{G} \mathbf{G}^{\textnormal{H}} \mathbf{\Theta}^{\textnormal{H}} \mathbf{h}_{{\tt r}, t}$ in \eqref{eq:radarSNR_MRT} is rewritten as $\tsum_{m=1}^{M} { \lambda_{\langle m \rangle} \big|\! \left[\mathbf{h}_{{\tt r}, t}^{\textnormal{H}} \mathbf{\Theta} \mathbf{U} \right]_m \!\big|^2 } $ and is upper bounded by $\tsum_{m=1}^{M} { \lambda_{\langle m \rangle} \left| \left[ \mathbf{h}_{{\tt r}, t} \right]_m \right|^2 }$ without the knowledge of $\mathbf{h}_{{\tt r}, t}$.
To reach this upper bound, the phase-shift matrix of the BD-RIS should be configured so that $\mathbf{\Theta} \mathbf{U} = \mathbf{I}_N$, which yields $\mathbf{\Theta} = \mathbf{U}^{\textnormal{H}}$ and, in turn, yields \eqref{eq:maxSNR}. Eventually, the radar beamforming that achieves \eqref{eq:maxSNR} is $\mathbf{W}_{\tt r} \triangleq \left[ \mathbf{w}_{{\tt r}, 1}, \mathbf{w}_{{\tt r}, 2}, \dots, \mathbf{w}_{{\tt r}, M} \right] = \frac{1}{\sqrt{M}} \mathbf{V}$ with the factor $\frac{1}{\sqrt{M}}$ ensuring ${\sum_{m=1}^{M}{\left\| \mathbf{w}_{{\tt r}, m} \right\|^2} = 1}$.
    This completes the proof of Theorem~\ref{theo:maxSNR}.
\end{IEEEproof}

As a direct corollary from the above proof, traditional diagonal RISs cannot realize maximum radar SNR. 
    This is because a diagonal RIS only applies independent phase shifts to each element along its diagonal. To achieve $\mathbf{\Theta} = \mathbf{U}^{\textnormal{H}}$, which involves off-diagonal couplings, a non-diagonal, fully connected RIS structure is required. 

\subsection{ZF-enabled Communication SINR}

We employ the ZF precoding technique to effectively suppress multi-user interference at each user ${k \in {\cal K}}$. 
Specifically, the beamforming vectors for the communication signals are designed as
    $\mathbf{W}_{\tt c} \!=\! \mathbf{H}_{\tt c} \left( \mathbf{H}_{\tt c}^{\textnormal{H}} \mathbf{H}_{\tt c} \right)^{-1} \! \mathbf{D}_{\tt c} \in \mathbb{C}^{M\times K}$, where 
    $\mathbf{W}_{\tt c} \triangleq \left[ \mathbf{w}_{{\tt c}, 1}, \mathbf{w}_{{\tt c}, 2}, \dots, \mathbf{w}_{{\tt c}, K} \right]$,
    $\mathbf{H}_{\tt c} \triangleq \left[ \mathbf{h}_1, \mathbf{h}_2, \dots, \mathbf{h}_K \right] \in \mathbb{C}^{M\times K}$,
    and $\mathbf{D}_{\tt c} \in \mathbb{R}^{K \times K}$ is a diagonal matrix ensuring the normalization ${\operatorname{Tr}\left( \mathbf{W}_{\tt c} \mathbf{W}_{\tt c}^{\textnormal{H}} \right) = 1}$. 
The $(k, k)^{\uth}$ entry of $\mathbf{D}_{\tt c}$ is \cite{Ammar2020WCL, Jiang2011TIT}
\begin{align}
\left[ \mathbf{D}_{\tt c} \right]_{k,k} 
    &=  \sqrt{ K^{-1} \left\| \mathbf{w}_{{\tt c}, k} \right\|^{-2} }
    =   \sqrt{ K^{-1} \left[ \left( \mathbf{H}_{\tt c}^\textnormal{H} \mathbf{H}_{\tt c} \right)^{-1} \right]_{k,k}^{-1} } \\
    &=  \sqrt{ K^{-1} \mathbf{h}_k^\textnormal{H} \mathbf{P}_{\mathbf{H}_k}^{\perp} \mathbf{h}_k },
\end{align}
where $\mathbf{H}_k \in \mathbb{C}^{M \times (K-1)}$ is a submatrix obtained by striking $\mathbf{h}_k$ out of $\mathbf{H}_{\tt c}$,
    $\mathbf{P}_{\mathbf{H}_k}^{\perp} \triangleq \mathbf{I}_{M} - \mathbf{H}_k \left( \mathbf{H}_k^\textnormal{H} \mathbf{H}_k \right)^{-1} \mathbf{H}_k^\textnormal{H}$ stands for the orthogonal projection onto the null space of $\mathbf{H}_k^\textnormal{H}$. 
    As a result, the communication SINR of user $k \in {\cal K}$ is expressed as
\begin{align}
\Gamma_{{\tt c},k}^{\textnormal{ZF}}
=   \frac{ 
        \frac{\bar{\gamma}_{{\tt c}, k}}{K} 
        \mathbf{h}_k^\textnormal{H} \mathbf{P}_{\mathbf{H}_k}^{\perp} \mathbf{h}_k
    }
    { 
        \frac{\bar{\gamma}_{{\tt r}, k}}{M} \left| \mathbf{h}_{k}^{\textnormal{H}} \mathbf{h}_{\tt r} \right|^2 + \left\| \mathbf{h}_{\tt r} \right\|^2
    } \left\| \mathbf{h}_{\tt r} \right\|^2,
\label{eq:sinr_zf_final}
\end{align}
where $\mathbf{h}_{\tt r} \triangleq \mathbf{h}_{{\tt r}, t}^\textnormal{H} \boldsymbol{\Theta} \mathbf{G}$ 
and $\bar{\gamma}_{{\tt r}, k} \triangleq \mathfrak{L}_{{\tt c}, k} P_{\tt r} / \sigma_k^2$.
    
\section{Distribution Functions of the Radar SNR and the Communication SINR}

In this section, we aim to derive closed-form expressions for the CDFs of the radar SNR and the communication SINR.
First, with adjustments to \cite[Eq. (38)]{Zanella2009TC}, we can rewrite the PDF of the unordered eigenvalues of the complex Hermitian matrix $\mathbf{G} \mathbf{G}^\textnormal{H}$ using the real-valued generalized hyper-Erlang distribution as
\begin{align}
f_{\lambda}(x)
    =   \sum_{m=2}^{2M} { \chi_m \frac{x^{N-M+m-2}}{(N-M+m-2)!} e^{-x} },~x > 0, 
\label{eq:lamdba}
\end{align}
where ${\chi_m \!\triangleq\! \frac{(-1)^m}{M L_{\textnormal{uc}}} (N \!-\! M \!+\! m \!-\! 2)!
    \sum_{q=1}^{m-1} \left| \mathbf{\Omega}^{(m-q, q)} \right|}$ with ${\sum_{m=2}^{2M}{\chi_m} \!=\! 1}$, 
${L_{\textnormal{uc}} \!\triangleq\! \prod_{m=1}^M (N \!-\! m)! (M \!-\! m)!}$ and the $(i,j)^{\uth}$ 
element of ${\mathbf{\Omega}^{(p, q)} \!\in\! \mathbb{R}^{(M-1)\times (M-1)}}$ is ${\big( \alpha_{i,j}^{(p, q)} \!+\! N \!-\! M \big)!}$ with $\alpha_{i,j}^{(p, q)}$ given in \cite[Eq. (37)]{Zanella2009TC}. 
It is noted that the unordered eigenvalues are statistically correlated but identically distributed.
    Moreover, based on \eqref{eq:lamdba}, we obtain 
    ${\mathbb{E}\left[ \lambda_m \right] \!= N}$~and
    $\mathbb{V}\left[ \lambda_m \right] \!= N M$, $\forall m \!\in {\cal M}$, based on 
    $\sum_{m=2}^{2M}{m\chi_m} = M+1$ and $\sum_{m=2}^{2M}{m^2\chi_m} = 1 - 3N + M + 2MN$.

\begin{Theorem}
\label{theo:pdfGammamax}
When $\frac{N}{M}$ is sufficiently large, the maximum radar SNR, $\Gamma^{\star}_{{\tt r}, t}$, can be characterized by the squared Erlang distribution with shape $M$ and scale $N$, with its CDF derived in a closed-form expression~as
\begin{align}
F_{\Gamma^{\star}_{{\tt r}, t}}(\gamma)
    \to   1 - e^{ -\frac{1}{N} \sqrt{\frac{M \gamma}{\bar{\gamma}_{{\tt r}, t}}} }
    \sum_{m=0}^{M-1} 
    \frac{1}{m!}
    \frac{\left( 
        \frac{M \gamma}{\bar{\gamma}_{{\tt r}, t}}
    \right)^\frac{m}{2}}{N^m},~\gamma > 0.
\label{eq:pdfGammamax}
\end{align}
\end{Theorem}

\begin{IEEEproof}
For convenience, let us denote $\gamma_{\tt r} \triangleq \mathbf{h}_{{\tt r}, t}^\textnormal{H} \mathbf{\Sigma} \mathbf{h}_{{\tt r}, t}$ and ${ \gamma_{{\tt r}, m} \triangleq \left| \left[\mathbf{h}_{{\tt r}, t} \right]_m \right|^2}$, where $\gamma_{{\tt r}, m}$ are statistically i.i.d.. We conclude that
${\gamma_{\tt r} = \sum_{m=1}^{M} \lambda_{\langle m \rangle} \gamma_{{\tt r}, m} =^d \sum_{m=1}^{M} \lambda_m \gamma_{{\tt r}, m}}$, i.e., the sum over all ordered eigenvalues of $\mathbf{G}\mathbf{G}^\textnormal{H}$ is statistically identical to the sum over all unordered eigenvalues. Moreover, since the mean and variance of each unordered eigenvalue are $N$ and $MN$, respectively, the variance of $\frac{\lambda_m}{N}$ decays at a rate of $\frac{N}{M}$ while its mean remains at 1. 
    Moreover, as the BD-RIS includes numerous RIS elements, often resulting in $\frac{N}{M}$ sufficiently large \cite{Liu2024WCL}, the BD-RIS renders the statistics of $\frac{\lambda_m}{N} N \gamma_{{\tt r}, m}$ predominantly governed by $N \gamma_{{\tt r}, m}$.
Moreover, since $\gamma_{{\tt r}, m}$ are i.i.d. unit-mean exponential RVs, the CDF of $N \gamma_{{\tt r}, m}$, denoted as $F_{N \gamma_t}(\gamma) $, corresponds to an exponential distribution~as ${F_{N \gamma_t}(\gamma) \!=\! 1 - e^{-\frac{\gamma}{N}}}$, for ${\gamma > 0}$. 
    Hence, $\gamma_{{\tt r}, t}$ is actually the sum of $M$ statistically i.i.d. exponential RVs and thus follows an Erlang distribution with shape $M$ and scale $N$. 
Specifically, the CDF of $\gamma_{{\tt r}, t}$ is expressed as
\begin{align}
F_{\gamma_{\tt r}}(\gamma)
    =   1 - e^{ -\frac{\gamma}{N} }
    \sum_{m=0}^{M-1}
    \frac{1}{m!} \left( \frac{\gamma}{N} \right)^m,~ \gamma > 0.
\label{eq:pdfGammamax}
\end{align}

Using the fact that for a positive continuous RV $X$ with CDF $F_X(x)$, we have $F_{a X^2}(x) = F_X\left( \sqrt{x/a} \right)$ for $a \in \mathbb{R}^+$, we obtain \eqref{eq:pdfGammamax}. This completes the proof of Theorem \ref{theo:pdfGammamax}.
\end{IEEEproof}

\begin{Theorem}
\label{theo:cdf_sinr_k}
When $\frac{N}{M}$ is sufficiently large, the closed-form expression for the CDF of the ZF-enabled communication SINR of the user $k \in {\cal K}$ is derived as
\begin{IEEEeqnarray}{rCl}
F_{\Gamma_{{\tt c}, k}^{\textnormal{ZF}}}(\gamma)
&&\to H\left( \varrho \textnormal{SIR}_k-\gamma \right) \! 
    \sum_{m=0}^{M-K} (-1)^m \xi_{K,m} \left( \frac{\gamma}{\varrho \textnormal{SIR}_k} \right)^{K+m-1}
\nonumber\\
&&\hspace{5pt}
    + H\left( \gamma-\varrho \textnormal{SIR}_k \right) \! 
    \sum_{m=0}^{M-1} (-1)^m \xi_{1,m} \! \left( \frac{\varrho \textnormal{SIR}_k}{\gamma} \right)^{m} \!\!\!,\!\!
\label{eq:cdf_sinr_k}
\end{IEEEeqnarray}
for $\gamma > 0$, where $H(\cdot)$ depicts the Heaviside unit step function,
    $\xi_{i, m} \!\triangleq\! \binom{M-i}{m} \frac{\textnormal{B}(K+m-1,M-K+i)}{\textnormal{B}(K-1,M-K+1)}$,
    $\textnormal{SIR}_k \!\triangleq\! \frac{P_{{\tt c}, k}}{P_{\tt r}}$ is the effective average signal-to-interference ratio (SIR) at user $k \!\in\! {\cal K}$, 
    $\varrho \!\triangleq\! \frac{N M}{K(M+N-1)}$,
    and $\textnormal{B}(n, m) \!=\! \frac{(n-1)! (m-1)!}{(n+m-1)!}$ is the beta function.
\end{Theorem}

\begin{IEEEproof}
For convenience, we denote in this proof $\mathbf{z}_{k}$ and $\mathbf{z}_{\tt r}$ as the first $M$ elements of $\mathbf{h}_{{\tt c}, k}$ and $\mathbf{h}_{{\tt r}, t}$, respectively.
    It is noted that $\frac{\mathbf{h}_k^\textnormal{H} \mathbf{P}_{\mathbf{H}_k}^{\perp} \mathbf{h}_k}{\left\| \mathbf{h}_k \right\|^2}
    \!=\! 1 \!-\! \frac{\mathbf{h}_k^\textnormal{H} \mathbf{H}_k \left( \mathbf{H}_k^\textnormal{H} \mathbf{H}_k \right)^{-1} \mathbf{H}_k^\textnormal{H} \mathbf{h}_k}{\left\| \mathbf{h}_k \right\|^2}$ which,~based on the analysis in \cite{Jiang2011TIT}, follows a beta distribution. 
Moreover, as a direct corollary from Theorem \ref{theo:pdfGammamax}, we have ${\left\| \mathbf{h}_k \right\|^2 \!\to^d\! N \left\| \mathbf{z}_k \right\|^2}$
when $\frac{N}{M}$ is sufficiently large. Hence, the ZF-enabled communication SINR of the user $k \in {\cal K}$ is characterized as
\begin{align}
\Gamma_{{\tt c}, k}^{\textnormal{ZF}}
\mathop{\to}\limits^d 
   \frac{
        \frac{N}{K} \bar{\gamma}_{{\tt c}, k} 
        \mathbf{z}_k^\textnormal{H} \mathbf{P}_{\mathbf{Z}_k}^{\perp} \mathbf{z}_k
    }
    {   
        \frac{M+N-1}{M} \bar{\gamma}_{{\tt r}, k}
        \frac{\left| \mathbf{z}_k^{\textnormal{H}} \mathbf{z}_{\tt r} \right|^2}{\left\| \mathbf{z}_{\tt r} \right\|^2} + 1
    }
\mathop{\to}\limits^d 
    \varrho \textnormal{SIR}_k
    \frac{
        \frac{\mathbf{z}_k^\textnormal{H} \mathbf{P}_{\mathbf{Z}_k}^{\perp} \mathbf{z}_k}{\left\| \mathbf{z}_k \right\|^2}
    }
    {   
        \frac{\left| \mathbf{z}_k^{\textnormal{H}} \mathbf{z}_{\tt r} \right|^2}{\left\| \mathbf{z}_{\tt r} \right\|^2 \left\| \mathbf{z}_k \right\|^2}
    },
\label{eq:SINR_MRT1}
\end{align}
where 
    $\mathbf{P}_{\mathbf{Z}_k}^{\perp} \!\triangleq\! \mathbf{I}_{M} \!-\! \mathbf{Z}_k \left( \mathbf{Z}_k^\textnormal{H} \mathbf{Z}_k \right)^{-1} \mathbf{Z}_k^\textnormal{H}$ is the orthogonal projection onto the null space of $\mathbf{Z}_k^\textnormal{H}$ with
    $\mathbf{Z}_k \in \mathbb{C}^{M \times (K-1)}$ denoting the submatrix obtained from $\mathbf{Z}_{\tt c} = [\mathbf{z}_1, \mathbf{z}_2, \dots, \mathbf{z}_K] \in \mathbb{C}^{M \times K}$ by removing $\mathbf{z}_k$, i.e., the $k^{\uth}$ column. 
Then, based on the analysis in \cite{Jiang2011TIT}, we find that 
    $\beta_{\tt r} \!\triangleq\! \frac{\left| \mathbf{z}_k^{\textnormal{H}} \mathbf{z}_{\tt r} \right|^2}{\left\| \mathbf{z}_{\tt r} \right\|^2 \left\| \mathbf{z}_k \right\|^2}$ and
    $\beta_k \!\triangleq\! \frac{\mathbf{z}_k^\textnormal{H} \mathbf{P}_{\mathbf{Z}_k}^{\perp} \mathbf{z}_k}{\left\| \mathbf{z}_k \right\|^2}$ are independent beta RVs with PDFs $f_{\beta_{\tt r}}(x) = (M-1) x^{M-2}$ 
    and $f_{\beta_k}(x) \!=\! \frac{(M-1)!}{(K-2)! (M-K)!} x^{K-2} (1-x)^{M-K}$, for $x \in [0,1]$, respectively. As a result, \eqref{eq:SINR_MRT1} becomes the ratio of two independent beta RVs with integer shapes. 
Specifically, 
\begin{align}
F_{\Gamma_{{\tt c}, k}^{\textnormal{ZF}}}(\gamma)
    &=  \Pr\left[ \varrho \textnormal{SIR}_k \beta_k \le \gamma \beta_{\tt r}, \beta_k \le 1, \beta_{\tt r} \le 1 \right] \nonumber \\
    &=  \Pr\left[ \beta_k \le \gamma \beta_{\tt r}, \beta_k \le \min\left( \gamma/\varrho \textnormal{SIR}_k, 1 \right) \right] \nonumber \\
    &=  \textstyle
    \int_0^{\min\left( \gamma/\varrho \textnormal{SIR}_k, 1 \right)}
    \bar{F}_{\beta_{\tt r}}\left( x/\gamma \right) f_{\beta_k}(x) \mathrm{d} x,
    \label{eq:integral_SINR}
\end{align}
where $\bar{F}_{\beta_{\tt r}}(x) = (1-x)^{M-1}$, for $x \in [0,1]$. 
    The integral is straightforward to solve and obtain \eqref{eq:cdf_sinr_k} since $\bar{F}_{\beta_{\tt r}}$ and $f_{\beta_k}$ only include power functions. Hence, we omit details here for brevity, completing the proof of Theorem~\ref{theo:cdf_sinr_k}.
\end{IEEEproof}

\section{Outage Probability and Network Parameter Estimation}

To effectively ensure the reliability of both sensing and communication, we define the following problem to estimate network parameters, represented by the vector $\mathbf{x} \triangleq \left[ x_1, x_2, \dots, x_L \right]^\textnormal{T}$, as
\begin{subequations}
\begin{align}
\mathcal{P}: &~\mathrm{find}~\mathbf{x} \triangleq \left[ x_1, x_2, \dots, x_L \right]^\textnormal{T} \in \mathbb{R}^L, \label{1a} \\
\mathrm{s}.\mathrm{t}.\ 
    & 	F(\mathbf{x}) 
        \triangleq 
        \max_{\forall k\in{\cal K}}~
        \left\{ 
            {\cal OP}_{{\tt c}, k},
            {\cal OP}_{{\tt r}, t}
        \right\} = {\cal OP}^{\textnormal{th}}, \label{1b} \\ 
    &   0 \le x_l \le b_l,~\forall l \in {\cal L} \triangleq \{1, 2, \dots, L\}, \label{1c}
\end{align}
\end{subequations}
where each $x_l$ is a parameter subject to an individual upper bound $b_l$,
	${\cal OP}_{{\tt c}, k}$ is the outage probability (OP) at user $k \in {\cal K}$, defined as the probability that the communication SINR drops below a given target threshold, denoted as $\bar{\gamma}_{{\tt c}, k}^\textnormal{th}$, 
	${\cal OP}_{{\tt r}, t}$ is the OP that the radar SNR drops below a given target threshold $\bar{\gamma}_{{\tt r}, t}^\textnormal{th}$,
	and ${\cal OP}^{\textnormal{th}}$ denotes the required maximum OP.
Mathematically speaking, the OPs are derived as
\begin{align}
{\cal OP}_{{\tt c}, k}
    &=\!  \Pr\!\Big(
        \Gamma_{{\tt c}, k}^{\textnormal{ZF}} \le \bar{\gamma}_{{\tt c}, k}^\textnormal{th}
    \Big)
    =   F_{\Gamma_{{\tt c}, k}^{\textnormal{ZF}}}\left( \bar{\gamma}_{{\tt c}, k}^\textnormal{th} \right), \\
{\cal OP}_{{\tt r}, t}
    &=\!  \Pr\!\Big(
        \Gamma^{\star}_{{\tt r}, t} \le \bar{\gamma}_{{\tt r}, t}^\textnormal{th}
    \Big)
    =   F_{\Gamma^{\star}_{{\tt r}, t}}\left( \bar{\gamma}_{{\tt r}, t}^\textnormal{th} \right),
\end{align}
for $k \in {\cal K}$.
    When $\mathcal{P}$ has a finite number of solutions, it is regarded as a discrete inverse problem, a parameter estimation problem, or a root-finding problem \cite{aster2018parameter}. In this section, we present the SNIS algorithm to obtain analytical solutions to the problem $\mathcal{P}$ in integral-form expressions. Numerical approaches are then discussed in Section \ref{sec:numerical_result}.
    Before diving into the algorithm, we provide insights into the OP. Based on Theorems 2 and 3, we describe the asymptotic behaviour of the outage probabilities in the following corollaries.

\begin{Corollary}
\label{cor:1}
When either $N \to \infty$ or $P_{\tt r} \to \infty$, we obtain the asymptotic expression of ${\cal OP}_{{\tt r}, t}$ as
\begin{align}
{\cal OP}_{{\tt r}, t}^\textnormal{asymp}
    =   \frac{1}{M!} \left( \frac{M }{\bar{\gamma}_{{\tt r}, t}} \bar{\gamma}_{{\tt r}, t}^\textnormal{th} \right)^\frac{M}{2} N^{-M},
\end{align}
which implies that the OP decays at a rate of $N^M$ as the number of BD-RIS elements increases and at a rate of $\frac{M}{2}$ as the total radar power budget $P_{\tt r}$ increases.
\end{Corollary}

\begin{Corollary}
\label{cor:2}
The asymptotic expression of ${\cal OP}_{{\tt c}, k}$, i.e., as $\textnormal{SIR}_k \to \infty$, is obtained as
\begin{align}
{\cal OP}_{{\tt c}, k}^\textnormal{asymp}
    =   \frac{\textnormal{B}(K-1,M)}{\textnormal{B}(K-1,M-K+1)}
    \left( \frac{\bar{\gamma}_{{\tt c}, k}^\textnormal{th}}{\varrho \textnormal{SIR}_k} \right)^{K-1},
\label{eq:OP_ck_aymp}
\end{align}
which implies that the OP at user $k \in {\cal K}$ decays at a rate of $K-1$ as $P_{{\tt c}, k}$ increases.
    However, due to interference from the radar signal, increasing the number of BD-RIS elements influences the decay rate until it approaches a limit, derived by plugging $\varrho \to \frac{M}{K}$ into \eqref{eq:OP_ck_aymp}, as $N \to \infty$. 
Beyond this point, adding more BD-RIS elements has no further effect on ${\cal OP}_{{\tt c}, k}$.
\end{Corollary}

\subsection{Mathematical Preliminary}

\begin{Lemma}
\label{lem:1}
{(\it Feasibility Condition)}
Let us define a uniformly-distributed random vector $\mathbf{X} = [X_1, X_2, \dots, X_L]^\textnormal{T}$ with statistically i.i.d. entries and $X_l \in [0, b_l]$, for $l \in {\cal L}$, the joint PDF of $\mathbf{X}$ and~$Y = F(\mathbf{X})$, denoted as $f_{\mathbf{X}, Y}\left(\mathbf{x}, y\right)$, is obtained~as
\begin{align}
f_{\mathbf{X}, Y}\left(\mathbf{x}, y\right)
    =   \delta\left( y - F(\mathbf{x}) \right) f_{\mathbf{X}}(\mathbf{x}),
\label{eq:joint_pdf_model_gaussian}
\end{align}
for $\mathbf{0} \preceq \mathbf{x} \preceq \mathbf{b}$ and $y \in \left( 0, 1 \right)$, where $\delta(x)$ is the Dirac's delta function.
Then, ${\cal P}$ is feasible if and only if there exists $\mathbf{0} \preceq \mathbf{x} \preceq \mathbf{b}$ satisfying $f_{\mathbf{X}, Y}\left( \mathbf{x}, y \right) > 0$. 
\end{Lemma}

\begin{IEEEproof}
Due to the total probability theorem, the joint PDF of $\mathbf{X}$ and $Y$ is determined as $f_{\mathbf{X}, Y}\left(\mathbf{x}, y\right) = f_{Y | \mathbf{X}}(y|\mathbf{x}) f_{\mathbf{X}}(\mathbf{x})$, 
    where $f_{\mathbf{X}}(\mathbf{x}) \!=\! \prod_{l=1}^L f_{X_l}(x_l)$, for $\mathbf{0} \!\preceq\! \mathbf{x} \!\preceq\! \mathbf{b}$, is the joint PDF of $\mathbf{X}$ and 
    $f_{Y | \mathbf{X}}(y|\mathbf{x})$ is the PDF of $Y$ given that $\mathbf{X} = \mathbf{x}$, is derived by taking the derivative of the corresponding conditional CDF $F_{\left. Y \right| \mathbf{X}}\left(\left. y \right| \mathbf{x} \right)$ with respect to $y$. By definition, we have
    $F_{\left. Y \right| \mathbf{X}}\left(\left. y \right| \mathbf{x} \right) = \Pr\left[ F(\mathbf{x}) \le y | \mathbf{X} = \mathbf{x} \right] =  H\left( y-F(\mathbf{x}) \right)$.
As a result, $f_{\left. Y \right| \mathbf{X}}\left(\left. y \right| \mathbf{x} \right) = \frac{\rm d}{{\rm d} y} H\left( y-F(\mathbf{x}) \right) = \delta(y-F(\mathbf{x}))$, which yields \eqref{eq:joint_pdf_model_gaussian}. This completes the proof of Lemma \ref{lem:1}.
\end{IEEEproof}

\begin{Corollary} 
\label{cor:pdfJoint}
Based on \eqref{eq:joint_pdf_model_gaussian}, the joint PDF of $\mathbf{X}_{(l)}$ and $Y$ is obtained by integrating $f_{\mathbf{X}, Y} (\mathbf{x}, y)$ over $\mathbf{x}^{(L-l)}$, which yields
\begin{align}
f_{\mathbf{X}_{(l)}, Y}\left( \mathbf{x}_{(l)}, y \right)
    &=\!\!
    \int_0^1 \!\!
    \cdots 
    \! \int_0^1 \!
    \delta\left( F\left( \mathbf{x}_{(l)}, \mathbf{P}^{(L-l)} \mathbf{u}^{(L-l)} \right) - y \right)
    \nonumber\\
    &\quad\times
    \left| \mathbf{B}_{(l)} \right|^{-1} \mathrm{d} \mathbf{u}^{(L-l)},~\forall l \in {\cal L},
\label{eq:pdfJoint}
\end{align}
for $\mathbf{0} \!\preceq\! \mathbf{x}_{(l)} \!\preceq\! \mathbf{b}_{(l)}$, $y \!\in\! (0,1)$, where ${\mathbf{B}_{(l)} \!\triangleq\!\operatorname{diag}\left( \mathbf{b}_{(l)} \right)}$ and $\mathbf{B} = \operatorname{diag}\left( \mathbf{b} \right)$.
    Here, $f_{\mathbf{X}_{(l)}, Y}\left( \mathbf{x}_{(l)}, y \right)$ denotes the likelihood of $\mathbf{x}_{(l)}$ to be the solution of $\mathcal{P}$.
In this context, ${\cal P}$ is infeasible if there exists $l \in {\cal L}$ satisfying $f_{\mathbf{X}_{(l)}, Y}\left( \mathbf{x}_{(l)}, y \right) = 0$, for all $\mathbf{0} \!\preceq\! \mathbf{x}_{(l)} \!\preceq\! \mathbf{b}_{(l)}$.
\end{Corollary}
\begin{Lemma}
\label{lem:2}
The CDF of the solution $X_l$, conditioned on $Y \!=\! y$ and ${\mathbf X}_{(l-1)} \!=\! {\mathbf x}_{(l-1)}$ is presented by \eqref{eq:cdfXnCond} at top of the next page, for $x_l \in [0, b_l]$, where $\mathbf{u} = [u_1, u_2, \dots, u_L]^\textnormal{T}$ is the integration variable.
\begin{table*}
\normalsize
\begin{align}
\normalsize
F_{X_l\left|\, \mathbf{X}_{(l-1)}, Y \right.}\left( x_l \left|\, \mathbf{x}_{(l-1)}, y \right. \right)
=   \frac{\left| \mathbf{B}_{(l)} \right|^{-1}}{f_{\mathbf{X}_{(l-1)}, Y}\left( \mathbf{x}_{(l-1)}, y \right)}
    \! \int_0^1 \!\!
    \cdots 
    \! \int_0^1 \!\!
    x_l \delta\left( F\left( \mathbf{x}_{(l-1)}, x_l u_l, \mathbf{B}^{(L-l)} \mathbf{u}^{(L-l)} \right) - y \right)
    \mathrm{d} \mathbf{u}^{(L-l+1)},
\label{eq:cdfXnCond}
\end{align}
\hrulefill
\vspace{-15pt}
\end{table*}
\end{Lemma}
\begin{IEEEproof}
Based on the total probability theorem, we derive the desired CDF as
\begin{IEEEeqnarray}{rCl}
G_{X_l}(x_l)  
&=& \textstyle 
    \int_0^{x_l} f_{X_l \left| \mathbf{X}_{(l-1)}, Y \right.}\left( x \left|\, \mathbf{x}_{(l-1)}, y \right. \right) \mathrm{d} x  \\
&=& \frac{ \int_0^{x_l} f_{\left. \mathbf{X}_{(t)} \right| Y}\left( \left. \mathbf{x}_{(l-1)}, x \right| y \right) \mathrm{d} x }
        { f_{\left.\mathbf{X}_{(l-1)}\right| Y}\left( \left. \mathbf{x}_{(l-1)} \right| y \right) } \\
&=& \frac{ x_l \int_{0}^{1} f_{\mathbf{X}_{(t)}, Y}\left( \mathbf{x}_{(l-1)}, x_l u, y \right) \mathrm{d} u }
        { f_{\mathbf{X}_{(l-1)}, Y}\left( \mathbf{x}_{(l-1)}, y \right) },
\end{IEEEeqnarray}
where we define $G_{X_l}(x_l) \triangleq F_{X_l \left| \mathbf{X}_{(l-1)}, Y \right.}\left( x_l \left|\, \mathbf{x}_{(l-1)}, y \right. \right)$ for convenience. 
    Plugging \eqref{eq:pdfJoint} into the last equality, we obtain \eqref{eq:cdfXnCond}. This completes the proof of Lemma \ref{lem:2}.
\end{IEEEproof}
\subsection{Successive Non-Inversion Sampling (SNIS)}

Since the denominator of \eqref{eq:cdfXnCond} is the joint PDF of $\mathbf{X}_{(l-1)}$ and $Y$, \eqref{eq:cdfXnCond} becomes undefined when $\mathbf{x}_{(l-1)}$ is not a solution of $\mathcal{P}$, as this results in $f_{\mathbf{X}_{(l-1)}, Y}\left( \mathbf{x}_{(l-1)}, y \right) = 0$.
    To prevent this joint PDF from vanishing, we propose utilizing the priory generated solutions $\mathbf{x}_{(l-1)}$ for generating $x_l$.
In order to generate $x_l$, we introduce the following Theorem.

\begin{Theorem}
\label{theo:4}
(Non-inversion sampling)
Given that ${Y \!=\! F(\mathbf{X})}$, each solution $X_l$, bounded within $[0, b_l]$ and conditioned on $Y \!=\! y$ and ${\mathbf X}_{(l-1)} \!=\! {\mathbf x}_{(l-1)}$, for $l \in {\cal L}$, is obtained as
\begin{align}
X_l = \int_0^1 b_l H\left( V - G_{X_l}(b_l u) \right) \mathrm{d} u,
    \label{eq:theo:4}
\end{align}
where $H(x)$ denotes the Heaviside unit step function and ${V \in [0, G_{X_l}(b_l)]}$ is a standard uniform sample. 
\end{Theorem}

\begin{IEEEproof}
By applying the change of variable $z \leftarrow G_{X_l}(b_l u)$, the right-hand side of \eqref{eq:theo:4} becomes $G^{-1}_{X_l}(V)$, which the inverse CDF of $X_l$ conditioned on $Y \!=\! y$ and ${\mathbf X}_{(l-1)} \!=\! {\mathbf x}_{(l-1)}$. 
    Hence, based on the inversion sampling, also referred to as inverse-transform-sampling method \cite{Deng2023SJ}, $G^{-1}_{X_l}(V)$ yields a replica identically distributed to $X_l$ for a uniform RV $V$.
This completes the proof of Theorem \ref{theo:4}. 
\end{IEEEproof}

\begin{algorithm}
\label{algo:SNIS}
  \caption{Successive non-inversion sampling}
  \KwIn{$F$, $b_l$ for $l=1,2,\dots,L$;}
  
  \For{$l=1$ \KwTo $L$}{
    Compute $f_{\mathbf{X}_{(l-1)}, Y}\left( \hat{\mathbf{x}}_{(t-1)}, y \right)$ using Corollary \ref{cor:pdfJoint}; \\
    \If{$f_{\mathbf{X}_{(l-1)},Y}\left( \hat{\mathbf{x}}_{(t-1)}, y \right) = 0$}{
        $\mathcal{P}$ is infeasible; \textbf{break};
    }
    Compute the conditional CDF of $X_l$ in Lemma \ref{lem:2}; \\
    Generate $\hat{x}_l$ based on Theorem \ref{theo:4}; \\
  }
  \KwOut{$\hat{\mathbf{x}} = \left( \hat{x}_1, \hat{x}_2, \dots, \hat{x}_L \right)$;}
\end{algorithm}

Accordingly, the proposed SNIS algorithm for finding a solution of the problem ${\cal P}$ is outlined in Algorithm \ref{algo:SNIS}.
    It is noted that the algorithm eliminates the need for deriving explicit inverse CDFs, thus offers a significant advantage over traditional inversion sampling that requires the inverse CDF, which is often impractical or impossible to derive. 

\section{Numerical Results}
\label{sec:numerical_result}

For practical purposes, we consider the 3GPP Urban Micro (UMi) model with unit antenna gains for the path losses \cite[Table B.1.2.1-1]{NOMA3GPP}, where $\mathfrak{L} = 10^{-2.27-2.6\log {f_c}|_{\textnormal{GHz}}}$ with a path loss fading exponent of ${\alpha \!= 3.67}$ and ${f_c \!= 2}$~GHz.
Here, we consider the BS is equipped with ${M = 4}$ antennas serving ${K = 3}$ users, uniformly distributed within a 2-dimensional (2D) network area of size $100 \times 100$ $\textnormal{m}^2$ with the BS being located at the origin. 
    Moreover, the target is at $(x_t, y_t) \!=\! (100, 0)$ $\textnormal{m}$, i.e., $100$ meters away from the BS, and the BD-RIS is at $(50, 50)$ $\textnormal{m}$. 
The RCS is set at ${\varsigma_{\tt r}^2 = 1}$ \cite{Liu2024WCL}, with a sensing threshold of $30$ dB and the noise power levels are set to $\sigma_k^2 = \sigma_{\tt r}^2 = -104$ dBm, $\forall k \in {\cal K}$. 
    Hereafter, we consider that $P_{{\tt c}, k} = P_{\tt c}$ and $\bar{\gamma}^{\textnormal{th}}_{{\tt c}, k} = \bar{\gamma}^{\textnormal{th}}_{\tt c} = 2^R-1$ with $R = 2$ bps/Hz, $\forall k \in {\cal K}$, which leads to ${\cal OP}_{{\tt c}, 1} = {\cal OP}_{{\tt c}, 2} = \cdots = {\cal OP}_{{\tt c}, K}$.

\begin{figure}[!h]
    \centering
    \includegraphics[width=0.9\linewidth]{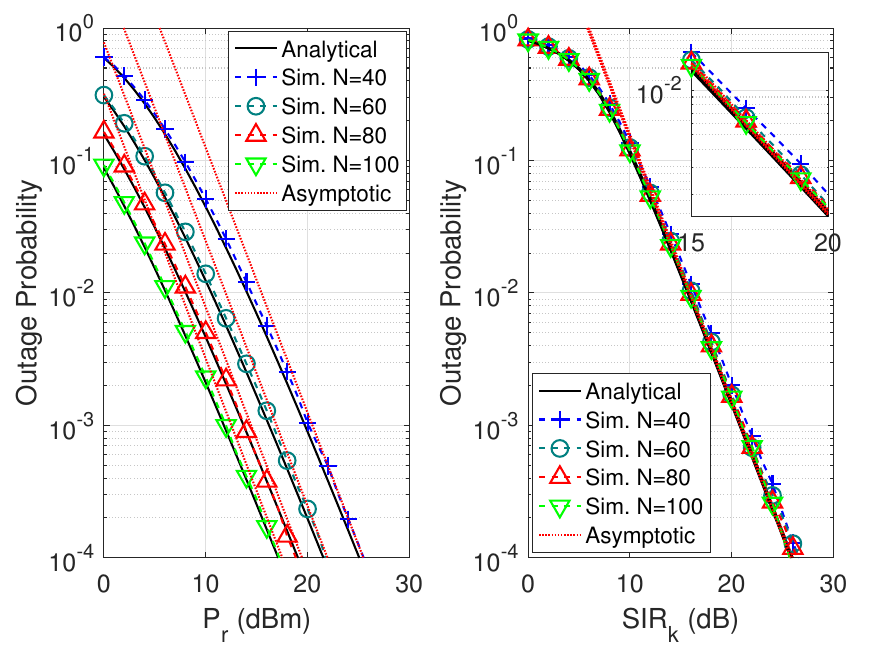}
    \caption{Radar OP (left) and communication OP (right) versus different number of BD-RIS elements.}
    \label{fig:1}
\end{figure}

In Fig. \ref{fig:1}, we presented analytical and simulated results for ${\cal OP}_{{\tt r}, t}$ and ${\cal OP}_{{\tt c}, k}$, $\forall k \in {\cal K}$. 
    As explained in Corollaries \ref{cor:1} and \ref{cor:2}, the radar and communication OPs decrease by $\frac{M P}{2}$ dB and $(K \!-\! 1) P$ dB, respectively, as the allocated power increases by $P$ dBm.
Moreover, Fig. \ref{fig:1} confirms that increasing the number of RIS elements significantly improves the radar OP while providing a negligible decrease in the communication OP, which is due to the interference from the radar signal as mentioned in Corollary \ref{cor:2}. 

Before presenting the results for the proposed SNIS algorithm, it's worth noting that numerical evaluation of integrals involving $\delta(x)$ in \eqref{eq:cdfXnCond} and \eqref{eq:pdfJoint} are quite challenging as $\delta(x)$ features an infinite peak at ${x=0}$ and is zero at all other points. To prevent numerical instability, we propose approximating Dirac's delta function~as 
\begin{align}
\delta(x) 
    \leftarrow \sigma_{\textnormal{err}}^{-1} \textnormal{K}\left( \sigma_{\textnormal{err}}^{-1} x \right) \triangleq \textnormal{K}_{\sigma_{\textnormal{err}}}(x),
\end{align}
where $\textnormal{K}$ is a Kernel function and ${\sigma_{\textnormal{err}} \!\ge\! 0}$ represents the trade-off between the model accuracy and numerical stability with ${\lim_{\sigma_{\textnormal{err}} \to 0} \textnormal{K}_{\sigma_{\textnormal{err}}}(x) = \delta(x)}$. 
    Hereafter, we adopt the sigmoid Kernel
    $\textnormal{K}(x) \!=\! \frac{\frac{2}{\pi}}{e^x+e^{-x}}$ 
    with $\sigma_{\textnormal{err}} \!=\! 10^{-4}$.
Moreover, the multiple integrals in \eqref{eq:cdfXnCond} and \eqref{eq:pdfJoint} can be numerically computed by adopting the $N_{\textnormal{MC}}$-point Monte Carlo integration method~as
\begin{align}
\int_0^1 \cdots \int_0^1 g(\mathbf{u}) \mathrm{d} \mathbf{u}
    \mathop{\to}\limits^{ N_{\textnormal{MC}} \to \infty }   
    \mathbb{E}\left[ g(\mathbf{U}) \right],
\label{eq:monte_integral}
\end{align}
where $g$ is the multivariate integrand and $\mathbf{U}$ is a standard uniform random vector with i.i.d. entries. Finally, \eqref{eq:theo:4} can be accurately evaluated using numerical integration methods, e.g., the $100$-point trapezoidal rule \cite{atkinson1991introduction}.

\begin{figure}[!h]
    \centering
    \includegraphics[width=0.9\linewidth]{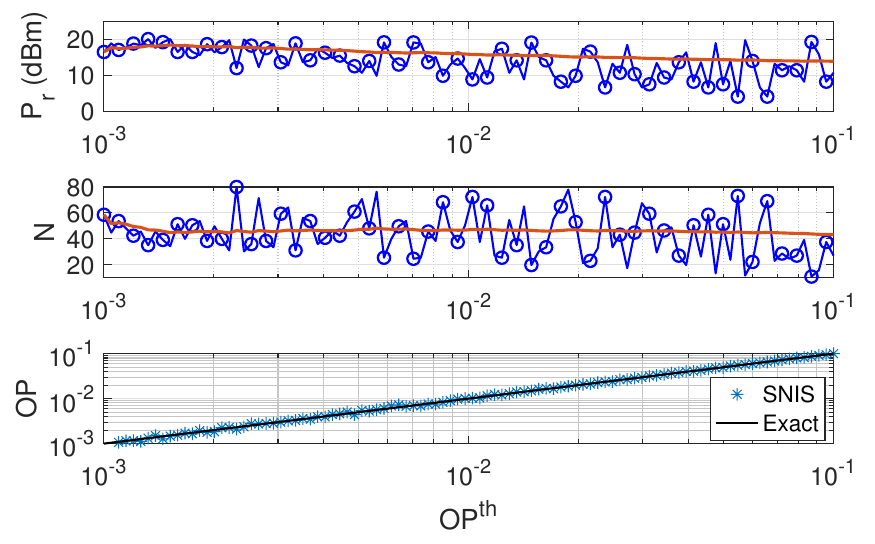}
    \caption{Allocated power for radar signals and number of BD-RIS elements as a function of the desired network outage levels, where $P_{\tt r} \le 20$ dBm, $N \le 80$ elements, and $P_{\tt c} = 15$ dBm, $\forall k \in {\cal K}$.}
    \label{fig:2}
\end{figure}

Fig. \ref{fig:2} presents numerical solutions of ${\cal P}$ by using the proposed SNIS algorithm with $T \!=\! 2$, where $F(\mathbf{x}) \!=\! F(P_{\tt r}, N^\ast)$, where $N \!\leftarrow\! \left\lceil N^\ast \right\rceil$. 
    The red curves show that as the desired network OP level increases, both $P_{\tt r}$ and $N$ decrease with less demanding network conditions. 
The observed fluctuations in the solutions are induced by three factors: i) the use of Monte Carlo integration in \eqref{eq:monte_integral}, ii) the generation of $\hat{x}$ in Step 7 due to $V$ being random, and iii) the fact that Fig. \ref{fig:2} presents only one solution among multiple feasible solutions of ${\cal P}$.
    As observed in the bottom figure, the SNIS algorithm produces results that accurately meet various target OP constraints.

\begin{figure}
    \centering
    \includegraphics[width = 0.9\linewidth]{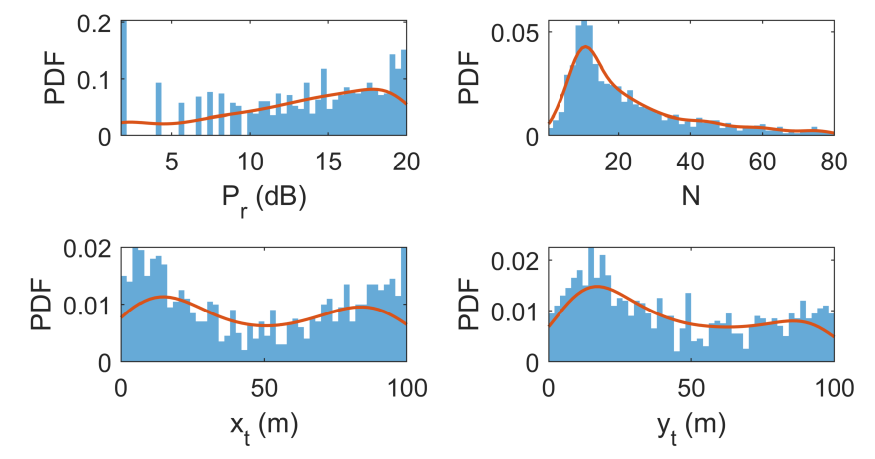}
    \caption{Marginal probability density function (PDF) of $P_{\tt r} \le 20$ dBm, $N \le 80$ element, $x_t \le 100$ m, and $y_t \le 100$ m with ${\cal OP}^\textnormal{th} = 10^{-2}$ and $P_{{\tt c},k } = 15$ dBm, $\forall k \in {\cal K}$.}
    \label{fig:3}
\end{figure}

Fig. \ref{fig:3} shows the marginal distribution of each solution of ${\cal P}$ with ${T \!=\! 4}$ using MATLAB's built-in function \texttt{histogram}, generated from $10^3$ random solutions. 
    When ${\cal OP}^\textnormal{th} = 10^{-2}$, $P_{\tt r}$ tends to approach its upper limit of $20$ dBm, requiring around $10$ BD-RIS elements.
Additionally, the target's position is less likely to be near the RIS, which is located at $(50, 50)$.

\section{Conclusion}
\label{sec:conclusion}

This paper presented a tractable approach to analyze the BD-RIS-assisted ISAC network's performance by developing closed-form expressions for the CDFs of the maximum radar SNR and ZF-enabled communication SINR. 
    A novel SNIS algorithm was proposed to tackle the network parameter estimation problem to satisfy various outage constraints.
Numerical results confirm the accuracy of our derived CDFs and showed that increasing BD-RIS elements improves radar performance but offers minimal gains in communication performance due to radar interference. Additionally, the SNIS algorithm provided accurate solutions to the network parameter estimation problem under various outage constraints.

\bibliographystyle{IEEEtran}
\bibliography{ref}

\end{document}